\documentclass{rrparticle}
\usepackage{graphicx}

\newcommand{\miktex}{\hbox{Mik\kern-.15em\TeX}}

\title{Collisional Dynamics of Solitons and Pattern Formation in an   Integrable  Cross Coupled Nonlinear Schr\"odinger equation with constant background} 
\author[1,a]{P.~S.~Vinayagam}
\author[1]{D.~Aravindha Krishnan}
\author[1]{R.~V.~Kamaleshwaran}
\author[2,b]{R.~Radha}
\affil[1]{Department of Physics, PSG College of Arts \& Science,
	Coimbatore-641 014, Tamilnadu, India.\newline
	Email:$^a${\em psvinayagam11@gmail.com}}
\affil[2]{Centre for Nonlinear Science (CeNSc), PG and Research Department of Physics, \newline
Government college for Women, Kumbakonam, Tamilnadu, India.\newline
	Email:$^b$ {\em vittal.cnls@gmail.com}}

\begin{document}
	\maketitle
	\begin{abstract}
		We investigate the dynamics arising out of the propagation of light pulses with different polarizations through a condensate (referred to as a constant background field) with cross coupling described by a coupled nonlinear Schr\"odinger equation(NLSE) type equation. We then employ Gauge and Darboux transformation approach to bring out the rich dynamics arising out of the background field and cross coupling. The collisional dynamics of bright solitons is found to be inelastic. The constant background field  is found to  facilitate   the periodic localization of light pulses during propagation. We have also unearthed breathers, bright-bright, bright-dark and dark-bright solitons of the coupled NLSE. While the amplitude of breathers oscillate with time as predicted, their maximum(or minimum) amplitude is found to remain a constant and the addition of cross coupling  only contributes to the rapid fluctuations in its amplitude over a period of time. In addition, the reinforcement of  cross coupling in the presence of constant wave field facilitates the interference of light pulses leading to interesting pattern formation among bright-bright, bright-dark and dark-bright solitons. The highlight of the results is that one obtains various localized excitations like breathers, bright and dark solitons by simply manipulating the amplitude of the constant wave field.
	\end{abstract}

\section{Introduction}
The exploration of nonlinear phenomena in physical systems has been a cornerstone of modern physics, revealing intricate behaviors that defy conventional understanding. To unearth  these unconventional phenomena, the nonlinear Schrödinger equation (NLSE) stands out as a fundamental mathematical framework for describing wave propagation in various fields, such as optics \cite{1,2} and plasma physics \cite{plas-3} to condensed matter systems \cite{ucgas-4} and beyond. However, when coupled with a background wave field that exhibits instability, the dynamics become significantly more complex, offering intriguing avenues for investigation \cite{5,6}.

Recently, the study of localized structures in various physical settings described by nonlinear partial differentials equations such as discrete Kadomtsev-Petviashvili (KP) equation \cite{r1}, nonlocal NLSE \cite{r2}, higher order integrable models and their bi-hamilton formulations \cite{r3}, fractional differential equations \cite{r4}, Parity-Time symmetric nonlocal equations \cite{r5}, Gross-Pitaevskii equation \cite{r6,akh3}, space shifted nonlocal NLSE \cite{r7}, and more \cite{r8,r9,r10,r11,r12,r13}  has attracted the attention due to their real world applications in the respective domains.

We explore  the intriguing realm of the coupled focusing nonlinear Schrödinger equation (CFNLSE) amidst an unstable background wave field. This scenario represents a fascinating intersection of nonlinear dynamics and instability, with profound implications across multiple disciplines \cite{mext}. Understanding the intricate interplay between nonlinear effects and unstable backgrounds is crucial for elucidating phenomena such as wave turbulence \cite{turb}, soliton dynamics \cite{sd}, and pattern formation \cite{sd}, all of which find applications in diverse fields including optical communications \cite{ha}, Bose-Einstein condensates \cite{be}, and plasma physics \cite{p}.

When coupled focusing NLSEs evolve in an unstable background wave field  like a condensate, the interplay between nonlinearity and instability introduces a myriad of intriguing phenomena. Instabilities can arise from various sources, including the presence of noise, external perturbations, or inherent instability in the system parameters. Understanding the dynamics of coupled NLSEs in such environments is crucial for predicting and controlling the behavior of nonlinear waves in practical applications.

In this paper, we focus on  the dynamics of  the propagation of  cross coupled light pulses through a condensate described by the  cross coupled  NLS type equation. We  transform it to an extended Manakov system through a similarity transformation and generate bright, dark solitons and breathers. We then bring out the impact of constant wave back ground (condensates) and cross coupling on the soliton dynamics.

The plan of the paper is as follows. In section II, we derive the mathematical (integrable) model governing the propagation of cross coupled light pulses through a constant background field. We then employ a similarity transformation to convert the cross coupled NLS type equation into an extended Manakov system in section II.  We then harness the Lax pair of the extended Manakov system to  generate soliton solutions employing both Gauge (vacuum seed) and Darboux transformation(nonzero seed) in section III . We then analyze the collisional dynamics of solitons under the combined impact of constant wave field and cross coupling. We also dwell upon the impact of cross coupling on breathers, bright-bright, bright-dark and dark-bright soliton solutions in the presence of constant background wave to bring about interesting pattern formation in Section IV. The results are then summarized at the end in section V.

\section{Model equation and Similarity transformation}
The dimensionless form of the coupled nonlinear Schr\"odinger equation (CNLSE) with a constant  background field and a cross  coupling is given by:
\begin{subequations}\label{moregeneral}
\begin{eqnarray}
	i q_{1t}(x,t)+\frac{1}{2}q_{1xx}(x,t)+ \left(|q_{1}(x,t)|^2+|q_{2}(x,t)|^2 - A^2 \right)q_{1}(x,t)+ \Omega q_{2}(x,t)&=&0 
	\notag\\ \\
	i q_{2t}(x,t)+\frac{1}{2}q_{2xx}(x,t)+ \left(|q_{1}(x,t)|^2+|q_{2}(x,t)|^2 - A^2 \right)q_{2}(x,t)+ \Omega q_{1}(x,t)&=&0 \notag
	\\
\end{eqnarray}
\end{subequations}

where \textit{x} and \textit{t} denote the spatial and temporal coordinates and $q_{1,2}$ represent the two complex field variables while \textit{A} represents the constant amplitude of the background field and $\Omega$ the cross coupling. The above model represents the propagation of two light pulses with different polarizations with cross coupling through a condensate (back ground field) and has been investigated recently \cite{mext} for ($\Omega=0$).  We plan to establish the  integrability of the  model with both constant background wave and cross coupling.\\

We now introduce the following similarity transformation \cite{similarity} to remove the cross coupling 
\begin{align}\label{treq}
\left(
\begin{array}{c}
	q_1(x,t) \\
	q_2(x,t) \\
\end{array}
\right)=\left(
\begin{array}{cc}
	a \cos (\Omega  t) & b \sin (\Omega  t) \\
	b \sin (\Omega  t) & a \cos (\Omega  t) \\
\end{array}
\right). \left(
\begin{array}{c}
	\psi _1(x,t) \\
	\psi _2(x,t) \\
\end{array}
\right)
\end{align}
where  the constants are  $a=1$  $\&$  $b=-i$  so that equation \eqref{moregeneral} gets transformed into an extended Manakov model  of the following form
\begin{subequations}\label{modelrabiabsent}
\begin{eqnarray}
	i \psi_{1t}(x,t)+\frac{1}{2}\psi_{1xx}(x,t)+ \left(|\psi_{1}(x,t)|^2+|\psi_{2}(x,t)|^2 - A^2 \right)\psi_{1}(x,t)&=&0\\
	i \psi_{2t}(x,t)+\frac{1}{2}\psi_{2xx}(x,t)+ \left(|\psi_{1}(x,t)|^2+|\psi_{2}(x,t)|^2 - A^2 \right)\psi_{2}(x,t)&=&0
\end{eqnarray}
\end{subequations}

In the next section, we exploit  the corresponding Lax pair of  the above equation \eqref{modelrabiabsent} and employ Gauge and Darboux Transformation (DT) using trivial and nontrivial seed solutions respectively to obtain   explicit soliton solutions.
\section{Lax pair, Gauge and Darboux transformations}
\subsection{Lax pair}
Equation  \eqref{modelrabiabsent} admits the following Lax pair \cite{mext}  which can be written in a compact form in terms of a pair of matrices as
\begin{subequations}
\begin{eqnarray}
	{\bf \Phi}_{X}&=&{\bf U  \Phi},  \label{Phix} \\
	{\bf \Phi}_{T}&=&{\bf V  \Phi},  \label{Phit}
\end{eqnarray}\label{laxcondition1}
\end{subequations}
where ${\bf U}$ and ${\bf V}$ are the 3$\times$3 matrices known as the Lax pair matrices, which assume the following  form:
\begin{align}
{\bf U}=\left(
\begin{array}{ccc}
	-i \lambda  & \psi _1(x,t) & \psi _2(x,t) \\
	-\psi _1{}^*(x,t) & i \lambda  & 0 \\
	-\psi _2{}^*(x,t) & 0 & i \lambda  \\
\end{array}
\right)
\end{align}
\begin{align}
{\bf V}=\left(
\begin{array}{ccc}
	V_{11} & V_{12} & V_{13} \\
	V_{21} & V_{22} & V_{23} \\
	V_{31} & V_{32} & V_{33} \\
\end{array}
\right)
\end{align}
\begin{align}
V_{11}&=i \lambda ^2-\frac{1}{2} i \left(\psi _1(x,t) \psi _1{}^*(x,t)+\psi _2(x,t) \psi
_2{}^*(x,t)-A^2\right) \nonumber\\
V_{12}&=i \lambda ^2-\frac{1}{2} i \psi _1(x,t) \psi_1{}^*(x,t)+\psi_2(x,t) \psi_2{}^*(x,t)-A^2 \nonumber\\
V_{13}&=-\lambda  \psi
_2(x,t)-\frac{1}{2} i \psi _{2x}(x,t) \nonumber\\
V_{21}&=\lambda \psi _1{}^*(x,t)-\frac{1}{2} i \psi _{1x}{}^*(x,t) \nonumber\\
V_{22}&=-\frac{1}{2} i \left(A^2-\psi _1(x,t) \psi _1{}^*(x,t)\right)-i \lambda ^2 \nonumber\\
V_{23}&=\frac{1}{2} i \psi_2(x,t) \psi_1{}^*(x,t) \nonumber\\
V_{31}&=\lambda \psi _2{}^*(x,t)-\frac{1}{2} i\psi _{2x}{}^*(x,t) \nonumber\\
V_{32}&=\frac{1}{2} i \psi _1(x,t) \psi _2{}^*(x,t) \nonumber \\
V_{33}&=-\frac{1}{2} i \left( A^2-\psi_2(x,t) \psi_2{}^*(x,t)\right)-i \lambda^2 \nonumber
\end{align}
where $\lambda$ is the complex spectral parameter. The consistency condition ${\bf \Phi}_{xt}={\bf \Phi}_{tx}$ leads to $ {\bf U}_{t}-{\bf V}_{x}+[{\bf U},{\bf V}]={\bf 0}$ which  generates the extended Manakov model given by eq. \eqref{modelrabiabsent}. 
\subsection{Gauge transformation and Soliton Solutions}
In this subsection, starting from a trivial seed solution,  we construct bright  soliton solutions and analyze the impact of constant wave field.

So, we start from the trivial seed as $\psi_{1}(x,t)=\psi_{2}(x,t)=0$ and employ Gauge transformation approach \cite{Llchaw}  to construct bright  soliton solutions of the following form

\begin{subequations}
\begin{eqnarray}\label{onesoliton}
	\psi_{1}(x,t)&=& -\frac{\epsilon _1  \exp \left[i \left(-2 \alpha  x+A^2 t+2
		(\alpha ^2 - \beta ^2) t+2 \chi _1\right)\right] \text{Sech}\left(2 \beta  x-4 \alpha  \beta  t-2 \delta _1\right)}{2 \sqrt{2}} \notag\\ \\
	\psi_{2}(x,t)&=&-\frac{\epsilon _2  \exp \left[i \left(-2 \alpha  x+A^2 t+2
		(\alpha ^2 - \beta ^2) t+2 \chi_1\right)\right] \text{Sech}\left(2 \beta  x-4 \alpha  \beta  t-2 \delta _1\right)}{2 \sqrt{2}} \notag\\
\end{eqnarray}
\end{subequations}
where, $\delta_1$ and $\chi_1$ are arbitrary real parameters, $\epsilon_1$ and $\epsilon _2$ are coupling constants which are complex in nature subject to the constraint 
\begin{align}
\beta=-\frac{\sqrt{\epsilon _1 \epsilon _{11}+\epsilon _2 \epsilon _{21}}}{4 \sqrt{2}}
\end{align}
where, $\epsilon _{11}$ and $\epsilon _{21}$ are conjugates of $\epsilon _{1}$, $\epsilon _{2}$ respectively with $\lambda$=$\alpha$+i$\beta$, where $\alpha$,and $\beta$ are arbitrary constants. Bright solitons of  the extended Manakov system with constant wave field  along with cross coupling given by eq. \eqref{moregeneral} can be straightforwardly written using the transformation given by eq.\eqref{treq}. From the above, we understand that one can not bring out  the impact of constant wave field on the density of the solitons driving the light pulses as it is present only in the phase term. 
The Gauge transformation approach \cite{Llchaw} can be
extended to generate multisoliton solutions. For example, the
two-soliton solution $\psi_{1,2}^{(2)}$ for the two modes can be expressed as
\begin{subequations}
\begin{align}
	\psi_{1}^{(2)} = 2 i \frac{A1}{B},\\
	\psi_{2}^{(2)} = 2 i \frac{A2}{B},
\end{align}\label{twosolsolution}
\end{subequations}
where
\begin{align}
A1&= \left(\lambda _2-\lambda _1\right) M_{121} M_{222} \left(\lambda _1-\mu _1\right) \left(\lambda _2-\mu
_2\right)+M_{122} M_{221} \left(\lambda _2-\mu _1\right) \left(\mu _2-\lambda _1\right) \left(\lambda _2-\mu
_2\right)\notag\\
&+\left(\mu _2-\mu _1\right) M_{111} M_{122} \left(\lambda _2-\mu _1\right) \left(\lambda _2-\mu
_2\right)+\left(\mu _2-\mu _1\right) M_{112} M_{121} \left(\lambda _1-\mu _1\right) \left(\mu _2-\lambda _1\right), \notag\\
A2&= \left(\lambda _2-\lambda _1\right) M_{112} M_{211} \left(\lambda _1-\mu _1\right) \left(\lambda _2-\mu
_2\right)+M_{111} M_{212} \left(\lambda _2-\mu _1\right) \left(\mu _2-\lambda _1\right) \left(\lambda _2-\mu
_2\right) \notag\\
&+\left(\mu _2-\mu _1\right) M_{212} M_{221} \left(\lambda _2-\mu _1\right) \left(\lambda _2-\mu
_2\right)+\left(\mu _2-\mu _1\right) M_{211} M_{222} \left(\lambda _1-\mu _1\right) \left(\mu _2-\lambda _1\right),\notag\\
B&=\left(M_{122} M_{211}+M_{121} M_{212}\right) \left(\lambda _1-\mu _1\right) \left(\lambda _2-\mu
_2\right)+\left(M_{112} M_{221}+M_{111} M_{222}\right) \left(\lambda _2-\mu _1\right) \left(\mu _2-\lambda
_1\right)\notag\\
&+\left(\lambda _2-\lambda _1\right) \left(\mu _2-\mu _1\right) \left(M_{111} M_{112}+M_{221}
M_{222}\right) \notag
\end{align}
with  $\lambda_j = \bar{\lambda_j}^* = \alpha_j + i \beta_j$ ,
\begin{align}
M_{11j}&= e^{-\theta_j}\sqrt{2};\quad\nonumber
M_{12j}=e^{-i\xi_j}\varepsilon_1^{(j)};\quad\nonumber
M_{13j}=e^{-i\xi_j}\varepsilon_2^{(j)};\nonumber\\
M_{21j}&= e^{i\xi_j}\varepsilon_1^{*(j)};\quad\nonumber
M_{22j}=e^{\theta_j}/\sqrt{2};\quad\nonumber
M_{23j}=0;\nonumber\\
M_{31j}&= e^{i\xi_j}\varepsilon_2^{*(j)};\quad\nonumber
M_{32j}=0;\quad\nonumber M_{33j}=e^{\theta_j}/\sqrt{2},\nonumber
\end{align}
where
\begin{align}
\theta _{j} &= 2  \beta _{j} x - 4\int (\alpha _{j}\beta_{j})dt+2\delta_{j},\nonumber\\
\xi _{j} &=A^{2} t- 2 \alpha _{j} x-
2\int(\alpha_{j}^{2}-\beta_{j}^{2})dt-2\chi _{j},\label{xij} \nonumber\\
\beta_j &=-\frac{\sqrt{\epsilon _{1j} \epsilon _{1j}^{\ast}+\epsilon _{2j} \epsilon _{2j}^{\ast}}}{4 \sqrt{2}}
\end{align}
where $j = 1, 2$ and, ${}^{\ast}$ describes the conjugate of the respective term.

\begin{figure}
\includegraphics[width=1.0\linewidth]{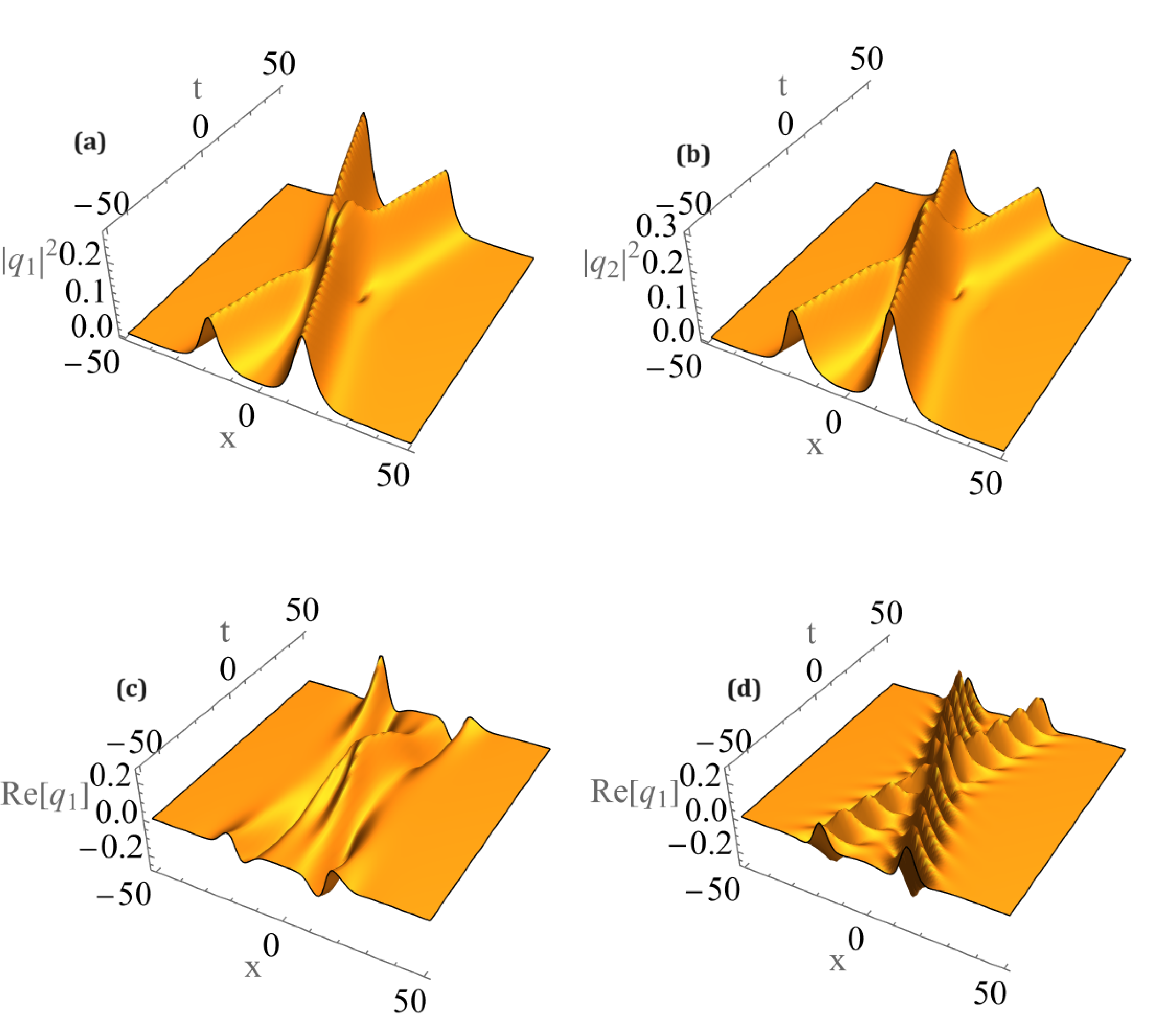}
\caption{Inelatic collision of light pulses   without cross coupling ($\Omega=0$) shown in panels (a,b) for the parametric choice $\alpha_1$=0.15, $\alpha_2$=-0.15, $\delta_i$=$\chi_i$=0.1 $\epsilon _{1j}$=0.5 where $j=1,2$ for $A=0$. Panel (c)  represents the real part of   $q_1$ for $A=0$ while panel (d) describes the  real part of   $q_1$ for non zero $A$}\label{collision}
\end{figure}

The  two soliton density profile is plotted in panels (a,b) of fig.\ref{collision}. From fig.\ref{collision} (a,b), we observe  that the  light pulses of the  extended Manakov model exchange energy among themselves undergoing inelastic collision \cite{psvcpb,inelastic} in the absence of constant wave field.  From fig.\ref{collision}(d), it is obvious that the constant wave field (or a condensate) interacts with light pulses thereby enhancing its amplitude at periodic intervals of time  while the light-matter interaction is missing in fig.\ref{collision}(c) for $A=0$. The behaviour of the constant wave field is similar to a reservoir feeding mater wave energy into the light pulses.

\begin{figure}
\includegraphics[width=1.0\linewidth]{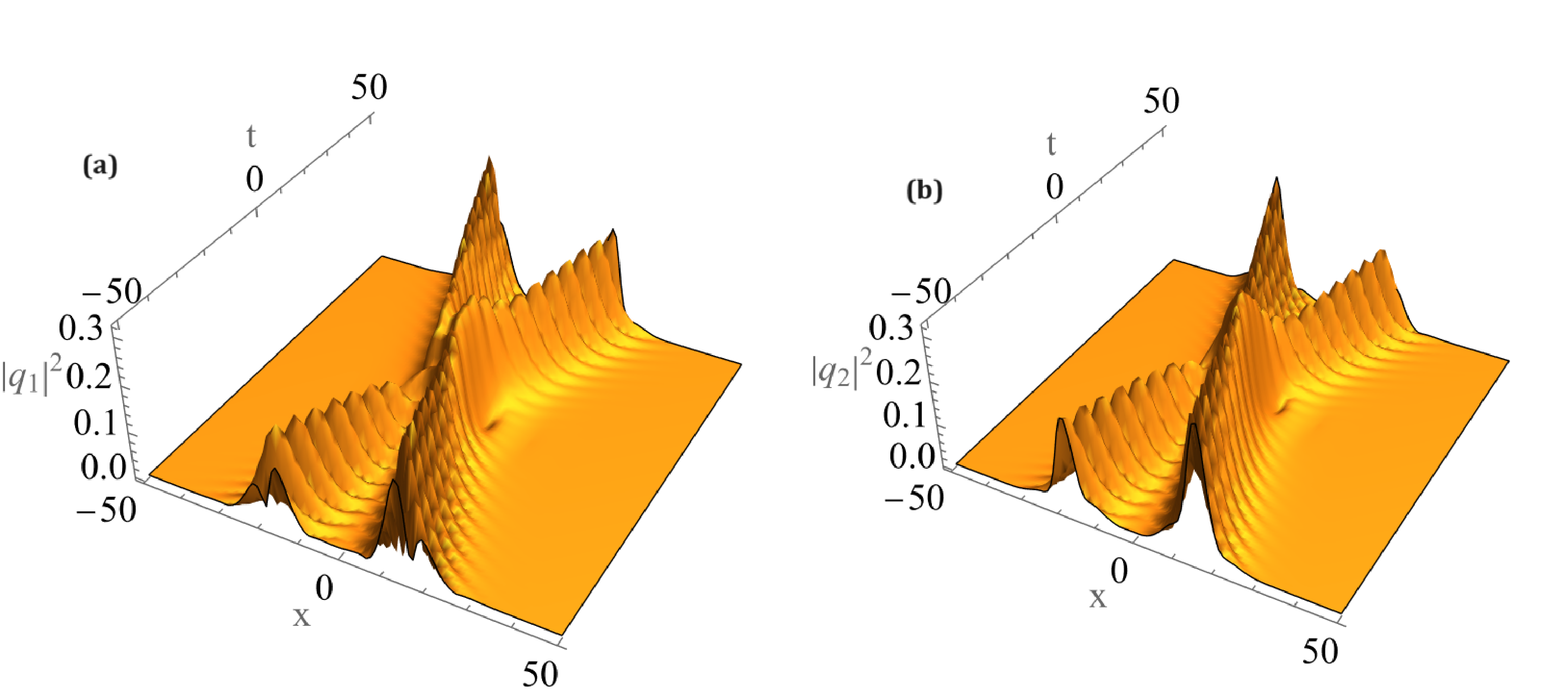}
\caption{Inelastic Bright-Bright solitons with oscillating amplitude for cross coupling parameter $\Omega= 0.5$ with the  rest of the parameters being the same as shown in fig.\ref{collision}}\label{rabigt}
\end{figure}

It is quite obvious that the addition of cross coupling only adds to the oscillation in the amplitude or the density of the solitons driving the light pulses as shown in fig.\ref{rabigt}.

\subsection{Darboux Transformation}
Since the impact of the constant wave field (or a condensate)  has been witnessed only in the real or imaginary  part of the wave function and not in the density of the  soliton solutions generated by means of Gauge transformation approach, it is better exploit an approach to generate the soliton solutions with a nontrivial seed solutions employing  "Darboux transformation" \cite{mateev}. We choose the following  non-trivial seed solution 
\begin{align}
\psi_{1}^{(0)} &= a e^{i A t} \;\;\& \;\; \psi_{2}^{(0)}= 0
\end{align}
which leads to the constraint relating  the amplitude of the solitons with the constant wave field  as 
\begin{align}
a=\sqrt{A^2+1}.
\end{align}
 With the above nontrivial  seed, the iteration employing Darboux transformation can be extended to generate soliton solutions of the   second order of the following form:
\begin{subequations}
\begin{align}
	\psi_{1}^{(2)}&=\psi_{1}^{(0)}+\frac{2 i \left| N_1\right| }{\left| R_2\right| }\\
	\psi_{1}^{(2)}&=\psi_{2}^{(0)}+\frac{2 i \left| N_2\right| }{\left| R_2\right| }
\end{align}\label{dttwosoltext}
\end{subequations}
where
\begin{align}
N_1&=\left(
\begin{array}{cccccc}
	\sqrt{A^2+1} \text{$\lambda _1$} e^{i t} & \sqrt{A^2+1} \text{$\lambda _2$} e^{i t} & \text{$\lambda _1^{\ast}$}
	\text{$\varphi_1^{\ast}$} & \text{$\lambda _2^{\ast}$} \text{$\varphi_2^{\ast}$} & 0 & 0 \\
	\sqrt{A^2+1} e^{i t} & \sqrt{A^2+1} e^{i t} & \text{$\varphi_1^{\ast}$} & \text{$\varphi_2^{\ast}$} & 0 & 0 \\
	\text{$\lambda _1$} \text{$\phi_1 $} & \text{$\lambda _2$} \text{$\phi_2 $} & 0 & 0 & \text{$\lambda _1^{\ast}$} \text{$\varphi_1^{\ast}
		$} & \text{$\lambda _2^{\ast}$} \text{$\varphi_2^{\ast}$} \\
	\text{$\phi_1 $} & \text{$\phi_2 $} & 0 & 0 & \text{$\varphi_1^{\ast}$} & \text{$\varphi_2^{\ast}$} \\
	\sqrt{A^2+1} \text{$\lambda _1$}^2 e^{i t} & \sqrt{A^2+1} \text{$\lambda _2$}^2 e^{i t} & \text{$\lambda _1$}^2
	\text{$\varphi_1^{\ast}$} & \text{$\lambda _2$}^2 \text{$\varphi_2^{\ast}$} & 0 & 0 \\
	\text{$\varphi_1 $} & \text{$\varphi_2 $} & -\sqrt{A^2+1} e^{-i t} & -\text{$\Psi_2^{\ast} $} & -\text{$\phi_1^{\ast} $} &
	-\text{$\phi_2^{\ast} $} \\
\end{array}
\right)\nonumber
\end{align}
\begin{align}
N_2&=\left(
\begin{array}{cccccc}
	\sqrt{A^2+1} \text{$\lambda _1$} e^{i t} & \sqrt{A^2+1} \text{$\lambda _2$} e^{i t} & \text{$\lambda _1^{\ast}$}
	\text{$\varphi_1^{\ast}$} & \text{$\lambda _2^{\ast}$} \text{$\varphi_2^{\ast}$} & 0 & 0 \\
	\sqrt{A^2+1} e^{i t} & \sqrt{A^2+1} e^{i t} & \text{$\varphi_1^{\ast}$} & \text{$\varphi_2^{\ast}$} & 0 & 0 \\
	\text{$\lambda _1$} \text{$\phi_1 $} & \text{$\lambda _2$} \text{$\phi_2 $} & 0 & 0 & \text{$\lambda _1^{\ast}$} \text{$\varphi_1^{\ast}
		$} & \text{$\lambda _2^{\ast}$} \text{$\varphi_2^{\ast}$} \\
	\text{$\phi_1 $} & \text{$\phi_2 $} & 0 & 0 & \text{$\varphi_1^{\ast}$} & \text{$\varphi_2^{\ast}$} \\
	\text{$\lambda _1$}^2 \text{$\phi_1 $} & \text{$\lambda _2$}^2 \text{$\phi_2 $} & 0 & 0 & \text{$\lambda _1$}^2
	\text{$\varphi_1^{\ast}$} & \text{$\lambda _2$}^2 \text{$\varphi_2^{\ast}$} \\
	\text{$\varphi_1 $} & \text{$\varphi_2 $} & -\sqrt{A^2+1} e^{-i t} & -\text{$\Psi_2^{\ast} $} & -\text{$\phi_1^{\ast} $} &
	-\text{$\phi_2^{\ast} $} \\
\end{array}
\right)\nonumber
\end{align}
\begin{align}
R_2&=\left(
\begin{array}{cccccc}
	\sqrt{A^2+1} \text{$\lambda _1$} e^{i t} & \sqrt{A^2+1} \text{$\lambda _2$} e^{i t} & \text{$\lambda _1^{\ast}$}
	\text{$\varphi_1^{\ast}$} & \text{$\lambda _2^{\ast}$} \text{$\varphi_2^{\ast}$} & 0 & 0 \\
	\sqrt{A^2+1} e^{i t} & \sqrt{A^2+1} e^{i t} & \text{$\varphi_1^{\ast}$} & \text{$\varphi_2^{\ast}$} & 0 & 0 \\
	\text{$\lambda _1$} \text{$\phi_1 $} & \text{$\lambda _2$} \text{$\phi_2 $} & 0 & 0 & \text{$\lambda _1^{\ast}$} \text{$\varphi_1^{\ast}
		$} & \text{$\lambda _2^{\ast}$} \text{$\varphi_2^{\ast}$} \\
	\text{$\phi_1 $} & \text{$\phi_2 $} & 0 & 0 & \text{$\varphi_1^{\ast}$} & \text{$\varphi_2^{\ast}$} \\
	\text{$\lambda _1$} \text{$\varphi_1 $} & \text{$\lambda _2$} \text{$\varphi_2 $} & -\sqrt{A^2+1} \text{$\lambda _1^{\ast}$}
	e^{-i t} & -\text{$\lambda _2^{\ast}$} \text{$\Psi_2^{\ast} $} & -\text{$\lambda _1^{\ast}$} \text{$\phi_1^{\ast} $} & -\text{$\lambda _2^{\ast}$}
	\text{$\phi_2^{\ast} $} \\
	\text{$\varphi_1 $} & \text{$\varphi_2 $} & -\sqrt{A^2+1} e^{-i t} & -\text{$\Psi_2^{\ast} $} & -\text{$\phi_1^{\ast} $} &
	-\text{$\phi_2^{\ast} $} \\
\end{array}
\right)\nonumber
\end{align}
\begin{align}
\Psi_{j}&=a \; \epsilon _{j} e^{\frac{1}{2} i t \left(A^2+2 \text{$\lambda_{j}$}^2\right)-i \text{$\lambda_{j}$} x}\nonumber\\
\phi_{j}&=b \; \epsilon _{j} e^{i \text{$\lambda_{j}$} x-\frac{1}{2} i t \left(A^2+2 \text{$\lambda_{j}$}^2\right)}\nonumber\\
\varphi_{j}&=d \; \epsilon _{j} e^{i \text{$\lambda_{j}$} x-\frac{1}{2} i t \left(A^2+2 \text{$\lambda_{j}$}^2\right)}\nonumber
\end{align}$j=1,2$ and, $^{\ast}$ describes the conjugate of respective term and $\lambda_j=\lambda_{jR}+ i \lambda_{jI}$.

\subsection{Host of Breather, Bright and Dark solutions}
Rewriting the above more general DT soliton solution given by Eq.\eqref{dttwosoltext} in a compact form, we arrive at the following expression:
\begin{align}
	\psi_{j}^{(2)}&= \alpha_{k}\Big[\frac{cosh(\lambda_1)cosh(B_1 -2i\lambda_2)+cosh(\lambda_2)cosh(B_2 -2i\lambda_1)}{cosh(\lambda_1)cosh(B_1)+cosh(\lambda_2)cosh(B_2)}\Big] \label{host}
\end{align}
where, 
\begin{align}
	B_1 &= X_1 + X_2 - \lambda_1+a_1, \notag \\
	B_2 &=-X_1 + X_2 + i \lambda_2 -i A \notag \\
	X_1 &= i \sqrt{J_1} Sinh(\lambda_1+i \lambda_2) \Big(x+\sqrt{J_1}Cosh(\lambda_1+i \lambda_2) t\Big) \epsilon_{1j}\notag\\
	X_2 &= -i \sqrt{J_1} Sinh(\lambda_1-i \lambda_2) \Big(-x+\sqrt{J_1}Cosh(\lambda_1-i \lambda_2) t\Big) \epsilon_{1j}\notag\\
	J_1 &= -(a \alpha_1^{2}+2 b \alpha_1 \alpha_2+d \alpha_2^{2}) \notag
\end{align}
where, $\alpha_{j}$, $\beta_{j}$, a, b, and d are arbitrary real parameters.\\

\begin{figure}
\includegraphics[width=1.0\linewidth]{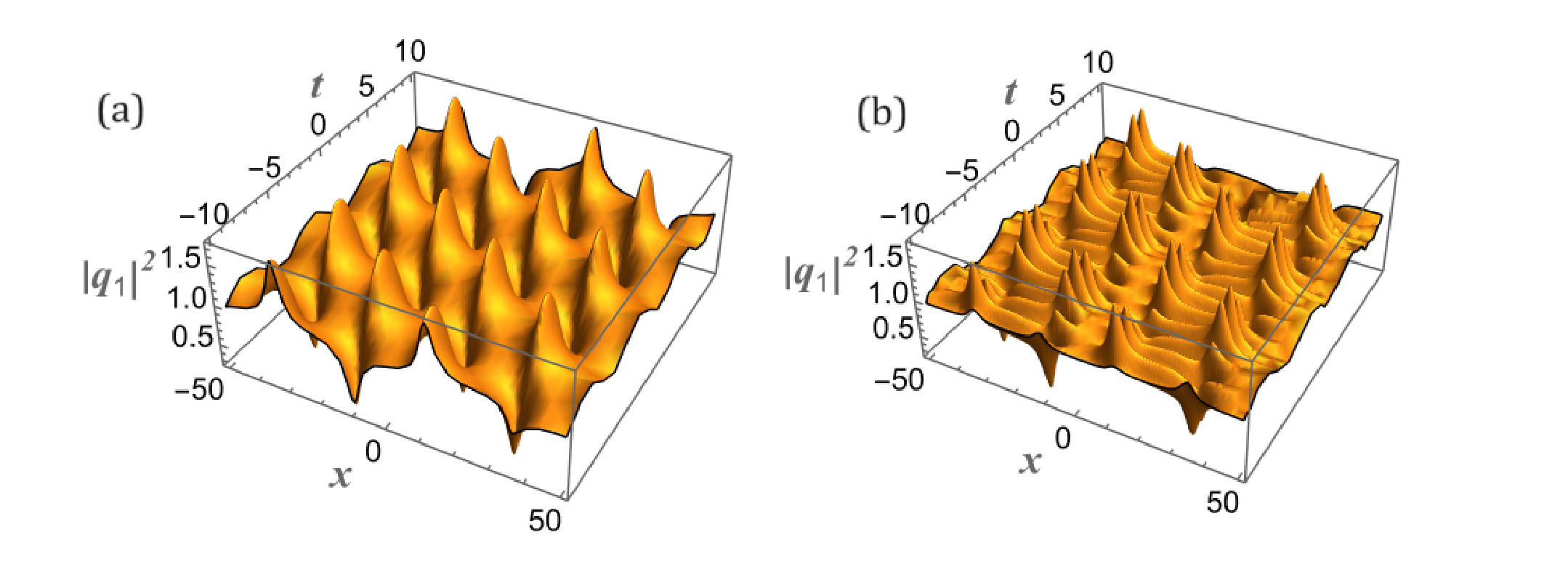}
\includegraphics[width=1.0\linewidth]{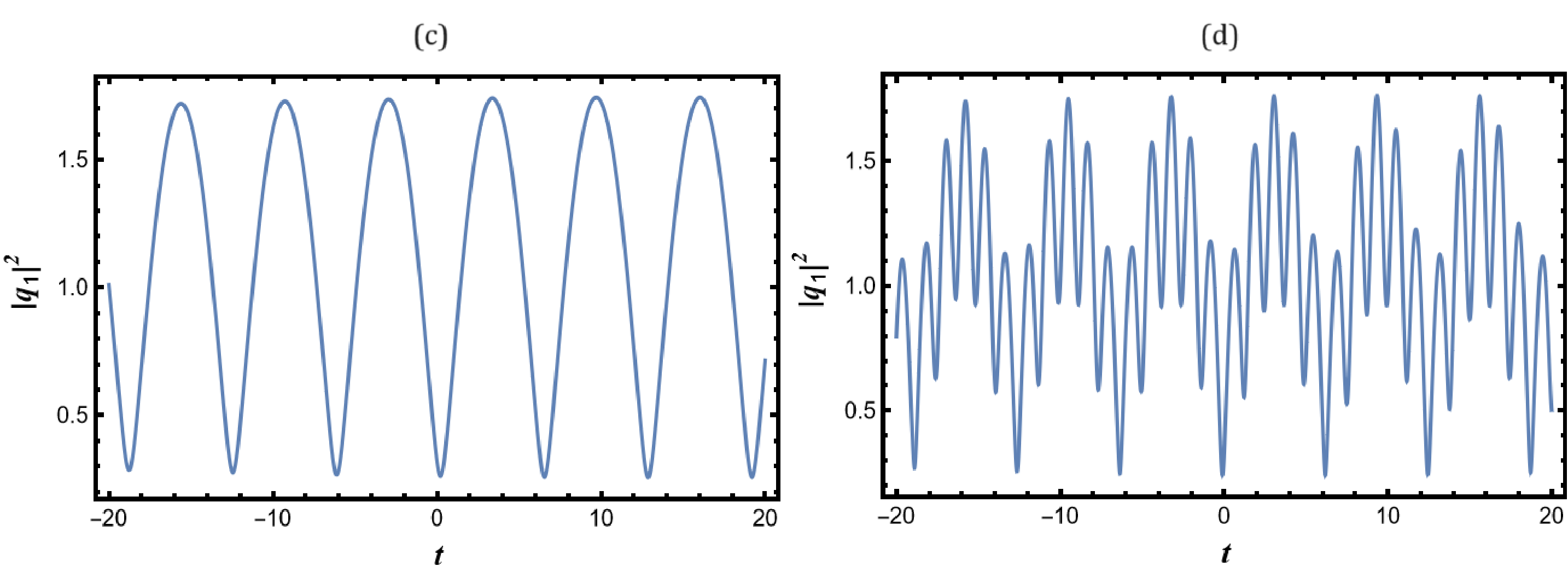}
\caption{Density profile of Breather solitons without $\Omega=0$ (a) and with cross coupling (b) $\Omega=0.5$  for the choice of parameters $\lambda_1$=0.008+0.07 $i$, $\lambda_2$=0.07+0.008 $i$, $\epsilon_{j}=1$, where $j=1,2$ and $A=0.05$. Panels (c) and (d) represent the corresponding time evolution of breathers shown in panels (a) and (b)  }\label{rabi_breather}
\end{figure}

\begin{figure}[h!]
\includegraphics[width=1.0\linewidth]{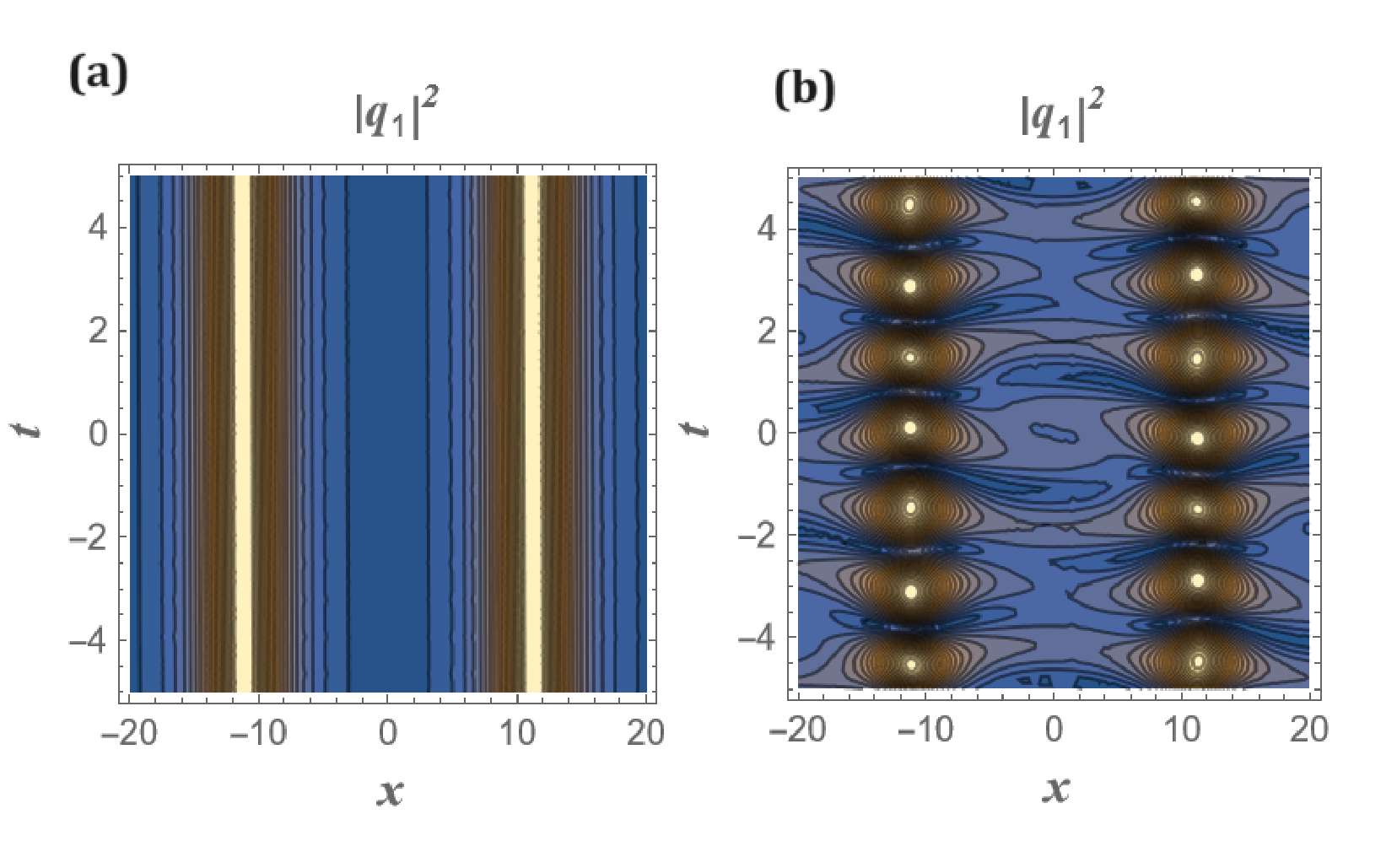}
\caption{Collisional dynamics of   bright-bright solitons  without cross ($\Omega=0$) (a) and with cross ($\Omega=0.5$) coupling  (b) for    $A= 0.5$ with the  other parameters being the   same as in fig.\ref{rabi_breather}}\label{rabi_cont_bb}
\end{figure}

\begin{figure}[h!]
\includegraphics[width=1.0\linewidth]{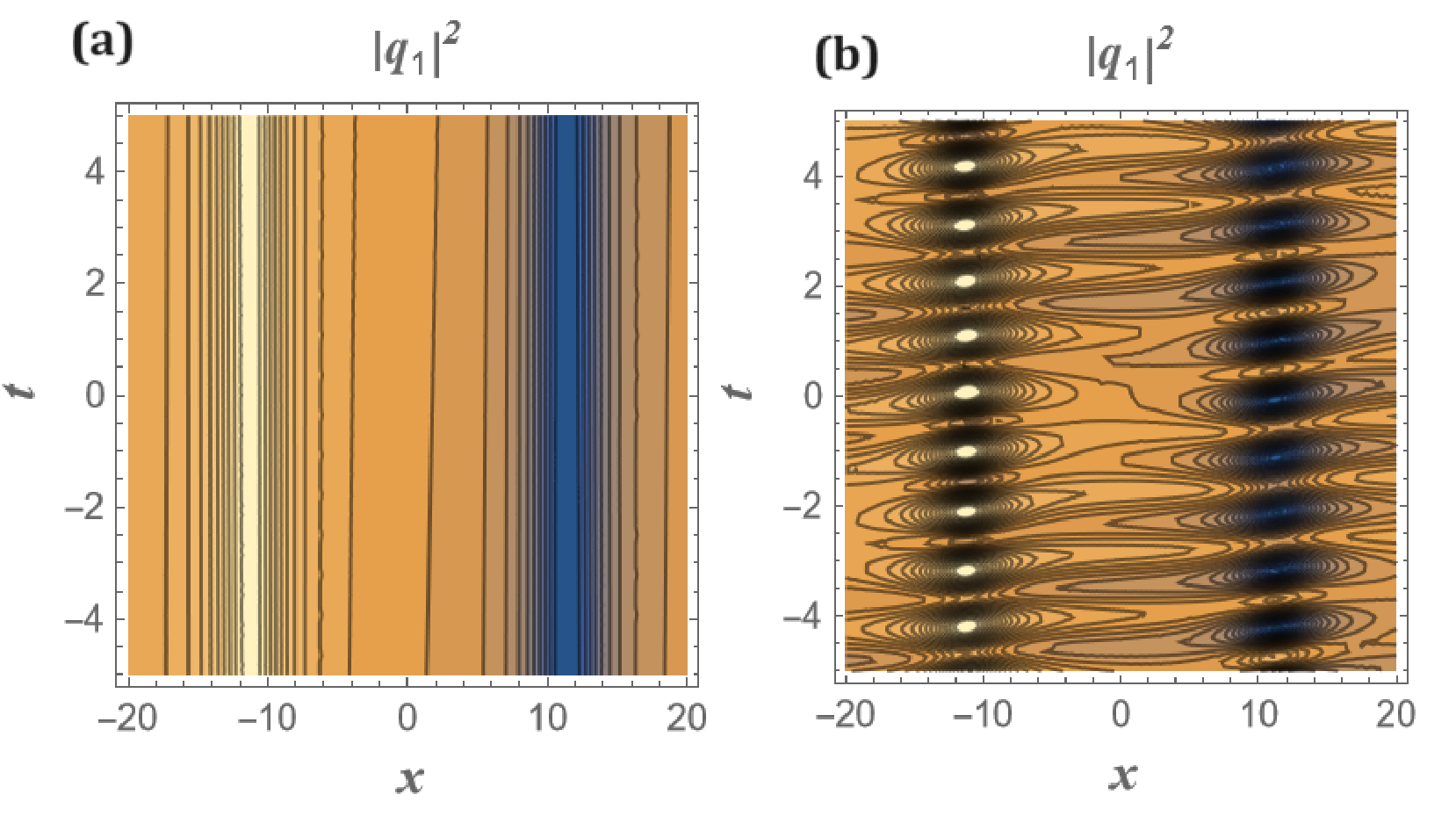}
\caption{Collisional dynamics of  bright-dark solitons without cross ($\Omega=0$) (a) and with cross ($\Omega=0.5$) coupling (b) for the choice of parameters $A= 0.9$ with  the other parameters being the  same as in fig.\ref{rabi_cont_bb}}\label{rabi_cont_bd}
\end{figure}

\begin{figure}[h!]
\includegraphics[width=1.0\linewidth]{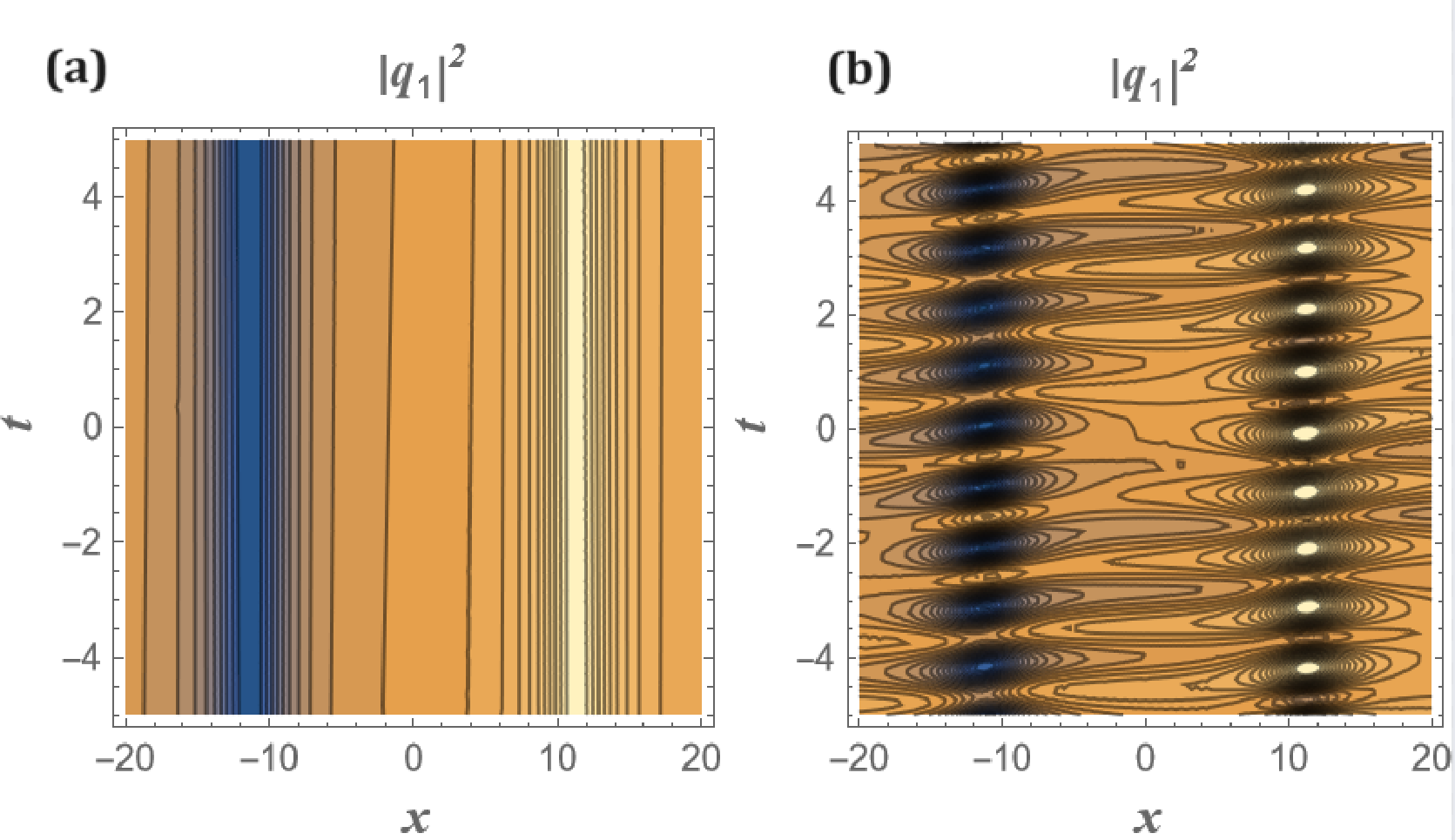}
\caption{Collisional dynamics  of dark-bright solitons without cross ($\Omega=0$) (a) and with cross ($\Omega=0.5$) coupling (b) for the choice of parameters $A= -0.9$ with the  other parameters being the  same as in fig.\ref{rabi_cont_bd}}\label{rabi_cont_db}
\end{figure}

A host of soliton pairs is associated with Eq.\eqref{host}. Upon examination, we observe that the soliton solutions comprise (i) a pair of solitons such as BB, BD, and DB, as well as (ii) an oscillating soliton wave. These solutions are all present on a constant background. We have presented an oscillatory solution for $J_1 <0$, which is the breather soliton \cite{akh1,akh2,akh3}, as illustrated in Fig.\ref{rabi_breather}, in addition to the pair of solitons (BB, BD, and DB) displayed in Figs.\ref{rabi_cont_bb}, \ref{rabi_cont_bd}, and \ref{rabi_cont_db}. A thorough explanation of how to generate these solutions is provided in the following section.

\subsection{Impact of Constant Wave field and Cross Coupling on the Soliton Dynamics}
As  discussed earlier, the impact of constant wave field on the  soliton solutions generated  by  Gauge transformation method is reflected only in the real or imaginary parts of the solution and not on the density of the solutions. To bring out the impact of constant wave field explicitly, Darboux tranformation \cite{mateev} has been employed to generate breathers, bright-bright, bright-dark and dark-bright solutions.

The density profile of breather solutions without  and with cross coupling is shown in panels (a)  and (b) of fig.\ref{rabi_breather}  while panels (c) and (d) show the corresponding time evolution of breathers. From fig.\ref{rabi_breather}, we understand that the amplitude (density) of the breather solitons  fluctuates with time with the minima and maxima remaining constant over a period of time. The addition of cross coupling only contributes to rapid oscillations in the amplitude of the breather solutions in a given cycle with the maxima and minima again remaining constant as shown in panel (d) of fig.\ref{rabi_breather}. The  breathers are found to be stable and are completely different from the one reported in ref\cite{mext}. In the case of bright-bright solitons, the presence of constant wave field (or a condensate) ensures that the amplitude of the solitons stays constant (shown in panel (a) of fig.\ref{rabi_cont_bb}) while the addition of cross coupling generates an interference pattern with bright fringes appearing at periodic intervals shown in panel (b) of fig.\ref{rabi_cont_bb}. This behaviour is repeated in the case of bright-dark and dark- bright soliton solutions shown in figs.\ref{rabi_cont_bd} and \ref{rabi_cont_db} where again the constant wave field ensures the amplitude of bright (dark) solitons remains constant while the reinforcement of cross coupling in addition to the constant wave field contributes to the fluctuations in the amplitudes where the density of bright (or dark) solitons becomes maximum at periodic intervals of time thereby generating interesting interference patterns in the process. The highlight of the results is that by simply manipulating the amplitude of the constant wave field, one obtains various localized excitations like breathers, bright and dark solitons and the two bound state counterparts.

\section{Conclusion}
In this paper, we have investigated the propagation of cross coupled light pulses through a constant wave field (or a condensate) described by a coupled NLS type equation and brought out the collisional dynamics of solitons. We have employed both Gauge and Darboux transformation approach to generate bright/dark solitons and breathers to bring out the impact of constant wave field and cross coupling. While cross coupling contributes to the fluctuations in the amplitude of the solutions, the constant wave field ensures that the density or the amplitude  of the solitons remains constant. The fact that the amplitude of the solitons driving the light pulses can be maintained constant by propagating through a condensate means that the results can have interesting ramifications in quantum information processing.

\section{Declaration of Competing Interest}
The authors declare that they  have no  competing financial interests or personal relationships that could have appeared to influence the work reported in this paper.
	
\section{Acknowledgments}
PSV wishes to express his deepest gratitude to the Principal and the Management of PSG College of Arts and Science for their moral support, encouragement throughout the tenure of this project. RR wishes thank DST-CRG (File number:CRG/2023/008153 dt. 30th January,2024) and DST-CURIE (DST/CURIE -PG/\\2022/54 dt 21/11/2022) for financial support.

%

\end{document}